\documentclass[12pt]{article}
\usepackage{amsmath,amssymb} 
\usepackage{subcaption}   
\usepackage{amsfonts}
\usepackage{enumerate}
\usepackage{ytableau}
\usepackage{hyperref}
\usepackage[
backend=biber,
style=numeric-comp,
sorting=none
]{biblatex}
\usepackage{tikz}
\usepackage{blkarray}  

\begin{document}

\def\im{\text{i}}
\def\eqa{\begin{eqnarray}}
\def\eqae{\end{eqnarray}}
\def\be{\begin{equation}}
\def\ee{\end{equation}}
\def\bea{\begin{eqnarray}}
\def\eea{\end{eqnarray}}
\def\ba{\begin{array}}
\def\ea{\end{array}}
\def\bd{\begin{displaymath}}
\def\ed{\end{displaymath}}
\def\eg{{\it e.g.~}}
\def\ie{{\it i.e.~}}
\def\Tr{{\rm Tr}}
\def\tr{{\rm tr}}
\def\>{\rangle}
\def\<{\langle}
\def\a{\alpha}
\def\b{\beta}
\def\c{\chi}
\def\del{\delta}
\def\e{\epsilon}
\def\f{\phi}
\def\vf{\varphi}
\def\tvf{\tilde{\varphi}}
\def\g{\gamma}
\def\h{\eta}
\def\j{\psi}
\def\k{\kappa}
\def\l{\lambda}
\def\m{\mu}
\def\n{\nu}
\def\w{\omega}
\def\p{\pi}
\def\q{\theta}
\def\r{\rho}
\def\s{\sigma}
\def\t{\tau}
\def\u{\upsilon}
\def\x{\xi}
\def\z{\zeta}
\def\D{\Delta}
\def\F{\Phi}
\def\G{\Gamma}
\def\J{\Psi}
\def\L{\Lambda}
\def\W{\Omega}
\def\P{\Pi}
\def\Q{\Theta}
\def\S{\Sigma}
\def\U{\Upsilon}
\def\X{\Xi}
\def\nab{\nabla}
\def\pa{\partial}
\newcommand{\lra}{\leftrightarrow}

\newcommand{\bc}{{\mathbb{C}}}
\newcommand{\br}{{\mathbb{R}}}
\newcommand{\bz}{{\mathbb{Z}}}
\newcommand{\bp}{{\mathbb{P}}}

\def\({\left(}
\def\){\right)}
\def\nn{\nonumber \\}

\newcommand{\red}{\textcolor[RGB]{255,0,0}}
\newcommand{\blue}{\textcolor[RGB]{0,0,255}}
\newcommand{\green}{\textcolor[RGB]{0,255,0}}
\newcommand{\cyan}{\textcolor[RGB]{0,255,255}}
\newcommand{\magenta}{\textcolor[RGB]{255,0,255}}
\newcommand{\yellow}{\textcolor[RGB]{255,255,0}}
\newcommand{\sky}{\textcolor[RGB]{135, 206, 235}}
\newcommand{\orange}{\textcolor[RGB]{255, 127, 0}}
\def\d{\operatorname{d}}
\def\ttbar{T$\overline{\text{T}}$ }
\def\arctan{\operatorname{actan}}
\def\arctanh{\operatorname{actanh}}

\title{\textbf{Thermodynamics and Holography of Three-dimensional Accelerating black holes}}
\vspace{14mm}
\author{Jia Tian$^{1,2}$\footnote{wukongjiaozi@ucas.ac.cn} and Tengzhou Lai$^{3}$\footnote{laitengzhou20@mails.ucas.ac.cn}}
\date{}
\maketitle

\begin{center}
	{\it $^1$State Key Laboratory of Quantum Optics and Quantum Optics Devices, Institute of Theoretical Physics, Shanxi University, Taiyuan 030006, P.~R.~China\\
 \vspace{2mm}
		$^2$Kavli Institute for Theoretical Sciences (KITS),\\
		University of Chinese Academy of Science, 100190 Beijing, P.~R.~China \\
  \vspace{2mm}
  $^3$School of Physical Sciences, University of Chinese Academy of Sciences, Zhongguancun East Road 80, Beijing 100190, P.~R.~China
	}
\vspace{10mm}
\end{center}

\begin{abstract}
We address the problem of describing the thermodynamics and holography of three-dimensional accelerating black holes. By embedding the solutions in the Chern-Simons formalism, we identify two distinct masses, each with its associated first law of thermodynamics. We also show that a boundary entropy should be included (or excluded) in the black hole entropy.
\end{abstract}


\section{Introduction}
Black holes play pivotal roles in general relativity, quantum gravity, and AdS/CFT holography \cite{Maldacena:1997re,Witten:1998qj,Gubser:1998bc}. An intriguing category is the accelerating black hole. Its exact solution existed solely in four-dimensional (4D) spacetime, termed the C-metric, until recently when 3D C-metrics describing compact accelerating black holes were introduced in \cite{Arenas-Henriquez:2022www,EslamPanah:2022ihg}.  4D C-metrics have been scrutinized in many aspects \cite{Letelier:1998rx,Bicak:1999sa,Podolsky:2000at,Pravda:2000zm,Dias:2002mi,Griffiths:2005qp,Krtous:2005ej,Dowker:1993bt,Emparan:1999wa,Emparan:1999fd,Emparan:2000fn,Gregory:2008br,Emparan:2020znc,Lu:2014sza,Ferrero:2020twa,Cassani:2021dwa,Ferrero:2021ovq,Appels:2016uha,Astorino:2016xiy,Appels:2017xoe,Anabalon:2018ydc,Anabalon:2018qfv,EslamPanah:2019szt,Gregory:2019dtq,Ball:2020vzo,Ball:2021xwt,Gregory:2020mmi,Kim:2023ncn,Clement:2023xvq,Hubeny:2009kz,Ferrero:2020laf,Ferrero:2021etw,Boido:2022iye,Griffiths:2006tk,Jafarzade:2017kin,Landgren:2024ccz}. However, the 3D C-metric which is supposed to be a simpler playground for  holographic exploration proves more enigmatic \cite{Astorino:2011mw,Xu:2011vp,Arenas-Henriquez:2022www,Arenas-Henriquez:2023hur,EslamPanah:2023ypz,EslamPanah:2023rqw,Cisterna:2023qhh}.

\begin{figure}[h]
    \centering
    \begin{minipage}[h]{0.48\linewidth}
        \includegraphics[scale=0.7]{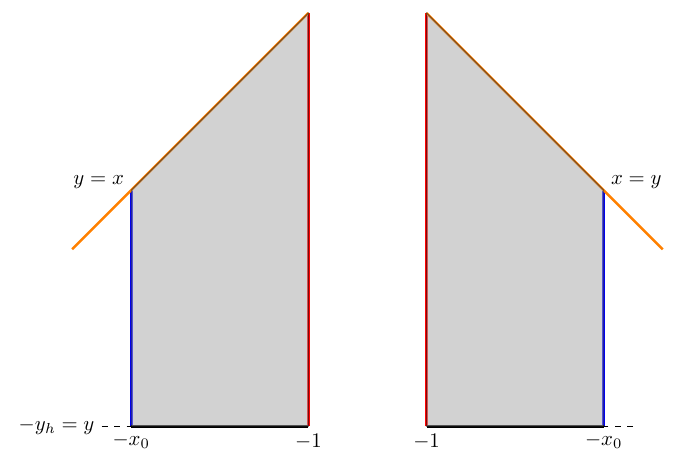}
        \caption{Two copies of the wedges.}
    \end{minipage}
    \hfill
    \begin{minipage}[h]{0.48\linewidth}
    \vspace{0.5cm}
        \includegraphics[scale=0.7]{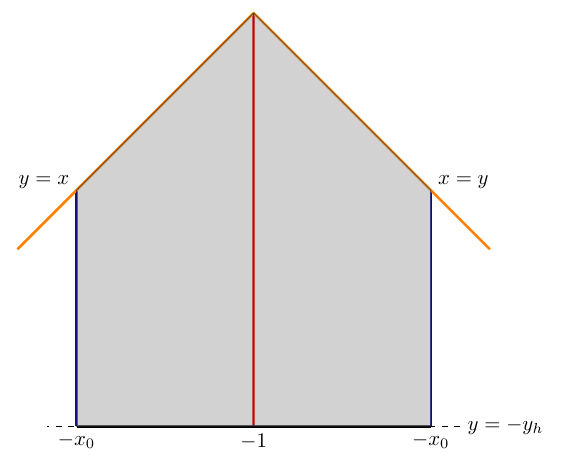}
        \caption{Gluing the EOW branes with the same color.}
    \end{minipage}
    \caption{Engineer a compact black hole by gluing EOW branes.}
    \label{glue}
\end{figure}

Three-dimensional C-metric has various phases describing  accelerating point particles or accelerating black holes. The metric takes the general form \cite{Arenas-Henriquez:2022www}:
\bea \label{eq:metric}
ds^2&=&\frac{1}{\Omega^2}\left[-P(y)a^2dt^2+\frac{dy^2}{P(y)}+\frac{dx^2}{Q(x)}\right],\\
\Omega &=& a(x-y),
\eea
where the free parameter $a$ is related to the acceleration and the AdS radius $\ell$ has been set to $1$.
We are interested in the accelerating black solution referred to as the Class II solution in \cite{Arenas-Henriquez:2022www} with the following metric functions
\bea 
&&Q(x)=x^2-1,\quad P(y)=\frac{1}{a^2}+1-y^2.
\eea 
Positivity condition of $Q(x)$ yields the following maximal range of coordinate $x$ is $ x>1 $ or $ x<-1$.
A 4D accelerating black hole is static due to the reality that it is anchored by a cosmic string. In three dimensions, it's the domain wall or the End-Of-the-World (EOW) brane, which serves as a fixed $x$ plane, fulfilling the cosmic string's function. Consequently, the solution can be incorporated into the AdS/BCFT formalism \cite{BCFT1,BCFT2}, such that the tension of the brane is dictated by the boundary equation of motion and Neumann boundary condition imposed on the EOW brane.
Gluing two copies of the wedge bounded by the EOW branes at $x=-1$ and $x=-x_0<-1$  as indicated in Fig.\ref{glue} results in a black hole geometry with compact spatial dimension.
We can introduce a new continuous angular variable as
\be 
x<-1,\quad x=-\cosh \psi,\quad \psi\in[-\psi_0,\psi_0]\,.
\ee
The "center" and event horizon of the black hole sits at $y=-\infty$ and $y=-y_h=-\sqrt{\frac{1+a^2}{a^2}}$. One possible way to diagnose the acceleration is to translate the C-metric into global AdS$_3$. We see that the "center" of the black hole corresponds to a point situated at a distance from the AdS$_3$'s center, indicating that the black hole is accelerated to counteract the negative pressure of the AdS space. In four dimensions, the acceleration is sustained by cosmic strings causing conical singularities along the string due to their back-reaction on the geometry. Notably, the black hole+string system possesses a well-defined thermodynamics \cite{Appels:2016uha}. In contrast, EOW branes do not back-react on the geometry rendering the 4D method of thermodynamics inapplicable. In AdS/BCFT, when a Ryo-Takayanaki (RT) surface intersects with EOW branes the associated entanglement entropy will receive a contribution from the EOW branes which is referred to as the boundary entropy. In the 3D metric case, the RT surface is the black hole horizon which suggests that we should consider the black hole+EOW system.

The primary distinction between the 3D and 4D C-metrics lies in the ambiguity of the black hole's (holographic) mass, which varies according to the chosen conformal structure \footnote{Actually it is very general in AdS space with odd dimensions, the holographic mass depends on the choice of conformal class \cite{Papadimitriou}.}.  The conformal structure dictates asymptotic boundary conditions. 

These novel feathers of 3D C-metric present challenges for formulating thermodynamics and holographic interpretation. Here we will address the problems in the Chern-Simons formalism of AdS gravity \cite{CS1,CS2} and the co-adjoint orbit theory \cite{orbit}. We begin by introducing a large family of static 3D solutions that are constructed in CS formalism and then clarify some subtleties about its thermodynamics and holographic interpretation. By embedding the 3D C-metric in this family, we derive the first law of thermodynamics and a Smarr formula. In the end, we will comment on the previous results \cite{Arenas-Henriquez:2023hur}, where a special conformal structure was chosen.
\section{Three-dimensional static solutions}
Since AdS$_3$ pure gravity lacks a local degree of freedom, its geometric configuration only depends on the asymptotic boundary condition. Utilizing the Chern-Simons formalism, numerous static solutions have been constructed \cite{Perez:2016vqo}. The metric for these solutions reads
\bea \label{eq:metric_static}
ds^2&=&\frac{dz^2}{z^2}+\frac{1}{z^2}(-f^2 dt^2+d\varphi^2)+\frac{u d\varphi^2+f(uf-2f'')dt^2}{2}\nonumber \\
&\quad &+z^2\frac{u^2d\varphi^2-(uf-2f'')^2dt^2}{16},
\eea 
where $f(\varphi),u(\varphi)$ are functions of the $\varphi$-circle $S^1$ and they are subject to the conditions
\be\label{eq:eom} 
\mathcal{D}f=0,\quad \mathcal{D}\equiv (\partial_\varphi u)+2u\partial_\varphi-2\partial^3_\varphi\,,
\ee 
which are equivalent to the Einstein's equation. The asymptotic boundary condition is reflected in the relation between $u$ and $f$ given by 
\be \label{eq:def_f}
f=2\pi \frac{\delta \mathcal{H}}{\delta u}.
\ee 
To support this boundary condition,  the Chern-Simons action needs to be complemented by the following
 the surface integral 
 \be \label{eq:boundary}
 B_{\infty}= \frac{\kappa}{8\pi}\int dt \, \mathcal{H}.
 \ee 
The metric functions $f(\varphi)$ and $u(\varphi)$ transform under an infinitesimal diffeomorphism $\epsilon(\varphi)$ as
\be
\delta f=\epsilon f'-f\epsilon',\quad\delta u=\mathcal{D}\epsilon. \label{eq:diff_u}
\ee 
The diffeomorphisms that are compatible with the boundary condition \eqref{eq:def_f} define the asymptotic symmetries. In particular, the charge which is associated with the time translation invariance is $\mathcal{H}$.
The transformation \eqref{eq:diff_u} implies that $f$ transforms like a vector in $\text{diff}(S^1)$ and $u$ transforms like a co-vector. If $\epsilon=f$ then $\delta f=\delta u=0$ thanks to \eqref{eq:eom}. By the co-adjoint orbit theory \cite{orbit},  the phase space of static solutions is described by the co-adjoint orbit $W_u=\frac{\text{diff}(S^1)}{f}$ with centralizer $f$.  In other words, if $u$ and $\tilde{u}$ belong to different orbits  they cannot be interchanged by diffeomorphisms. To have a smooth geometry, we will assume that $f$ is positive definite. Thus in each orbit, We can always find a representative $\varphi_0\sim \varphi_0+2\pi$ such that $f(\varphi_0)=f_0$ is constant. Other representations $\varphi$ are related to $\varphi_0$ via
\bea 
\oint\frac{d\varphi_0}{f_0}=\oint\frac{d\varphi}{f(\varphi)},
\eea 
where $L_\varphi$ is the circumference  of the $\varphi$-circle. When $f$ is transformed into a constant, simultaneously $u$ becomes a constant due to the invariant combination
\be \label{eq:cons}
uf^2-2ff''+(f')^2\equiv u_0f_0^2,
\ee
under a diffeomorphism. The invariant scalar product between a vector $f$ and a co-vector $u$ can be defined as
\be 
\oint d\varphi \, fu-2\oint f_0\{\varphi,\varphi_0\}d\varphi_0=2\pi  u_0f_0,
\ee 
where $\{\varphi,\varphi_0\}$ is the Schwarzian derivative.

To sum up, the phase space of the static solutions described by the metric \eqref{eq:metric_static} is the co-adjoint orbit $W_\mu$ which is labeled  by two constant $f_0$ and $u_0$ linked by \eqref{eq:def_f}. 

\subsection{Thermodynamics and holography}
The metric \eqref{eq:metric_static} is already written in the Fefferman-Graham (FG) gauge \cite{graham1985charles,Fefferman:2007rka} such that the holographic stress-energy tensor $T_{\mu\nu}$ \cite{Balasubramanian:1999re} can be easily read off. Integrating the component $T_{tt}$ along the $S^1$ gives the holographic mass
\be \label{eq:adm}
M_{\text{holo}}[\varphi]=\frac{1}{16\pi G_N}\oint d\varphi fu=\frac{c}{12}u_0f_0+\frac{c}{12\pi}f_0\oint\{\varphi,\varphi_0\}d\varphi_0,
\ee 
where the Brown-Henneaux relation $c=3/(2G_N)$ is used\cite{Brown:1986nw}. The metric \eqref{eq:metric_static} has an event horizon at
\be 
z_h=\sqrt{\frac{4f}{uf-2f''}},
\ee 
which is associated with a Hawking temperature and a Bekenstein-Hawking entropy \cite{Dymarsky:2020tjh}:
\bea\label{eq:tem}
T_{H}=\frac{\sqrt{u_0}f_0}{2\pi},\quad S_{BH}=\frac{\pi c}{3}\sqrt{u_0}\,.
\eea 
It may be a little surprising that holographic mass $M_{\text{holo}}$ depends on the orbit representative while the black hole entropy $S_{BH}$ does not. This is due simply to the fact  that the (asymptotic) charges only act on the asymptotic boundary but not on the horizon. It is also possible to define surface charges near the horizon \cite{Grumiller:2019fmp} by imposing a boundary condition to the horizon instead of to the conformal boundary. They correspond to soft modes \cite{soft} which contribute to the black hole entropy but not to the  mass. 

Now we are ready to derive the first law by varying the solution \eqref{eq:metric_static}. The variation can be split into two directions,  along the orbit and transverse to it.
The variation of $M_{\text{holo}}$ has both of these two contributions:
\bea 
\delta M_{\text{holo}}[\varphi]&=&\frac{c}{12}\(f_0\delta u_0+(u_0+\frac{1}{\pi}\oint\{\varphi,\varphi_0\})d\varphi_0)\delta f_0\) \nonumber \\
&+& \frac{c}{12\pi}f_0\oint \delta\(\{\varphi,\varphi_0\}\)d\varphi_0 .
\eea  
However, the variation of entropy is simply
\be 
T_{H}\delta S_{BH}=\frac{c}{12} f_0\delta u_0.
\ee 
It suggests the first law should be modified as
\be 
\delta M_{\text{holo}}[\varphi]=T_{H}\delta S_{BH}+\rho_{\varphi}\delta f_0+\oint\Omega_\varphi\delta\varphi \, d\varphi_0,
\ee 
where we have defined
\bea 
\rho_\varphi &=&\frac{c}{12}(u_0+\frac{1}{\pi}\oint d\varphi_0\{\varphi,\varphi_0\}),\\
\Omega_\varphi &=&\frac{c}{12\pi}\(\frac{\varphi''^2-\varphi'''\varphi'}{{\varphi'}^3}\)'.
\eea 
The variation $\delta f_0$ corresponds to the change of the proper length of the $S^1$ circle so $\rho_\varphi$ can be interpreted as the energy density.
It is easy to show that $\Omega_\varphi=0$ when $\varphi=c_1+c_2 \tanh{c_s(\varphi_0+c_4)}\equiv\varphi_s$ or $\varphi=\varphi_0$ and in these two cases the first law reduces to the standard one
\be 
\delta{M}_{\text{holo}}[\varphi_{s,0}]=T_{H}\delta S_{BH}+\rho_{\varphi_{s,0}}\delta f_0,
\ee 
with $\rho_{\varphi_s}=c(u_0-2c_s^2/\pi)/12$ and $\rho_0=cu_0/12$.

Alternatively, we can interpret the failure of first law $\delta M_{\text{holo}}\neq T_{H}\delta S_{BH} $ as a mismatch between the holographic mass and the thermodynamics mass. Examining the  asymptotic symmetry, the energy defined by asymptotic time translation invariance is $M_{\text{th}}\equiv \frac{c}{12}\mathcal{H}$ which is invariant under all the diffeomorphism compatible with the boundary conditions and thus the free energy of both the black hole and its dual boundary CFT should be \cite{Perez:2016vqo,Dymarsky:2020tjh}:
\be 
\mathcal{F}_{\text{free}}=M_{\text{th}}-T_HS_{BH},
\ee 
and the first law 
\be \label{eq:firstlaw}
\delta M_{\text{th}}=T_H\delta S_{BH},
\ee 
directly follows from \eqref{eq:def_f}. The Smarr formula also takes the standard form:
\be 
2M_{\text{th}}=T_{H}S_{BH}.
\ee 
The distinction between the holographic mass $M_{\text{holo}}$ and the thermodynamic mass $M_{\text{th}}$ originates from that the action of the bulk gravity theory is not simply given by standard AdS action:
\be \label{eq:onshell}
I_{\text{AdS}}=-\frac{1}{16\pi G_N}\int \sqrt{g}(R+2)-\frac{1}{8\pi G_N}\int \sqrt{h}(K-1).
\ee 
Indeed, substituting the Euclidean version of the metric \eqref{eq:metric_static} into \eqref{eq:onshell} leads to the on-shell action
\be 
I_{\text{on-shell}}=\beta(M_{\text{holo}}-T_HS_{BH})\neq \mathcal{F}_{\text{free}}.
\ee 
Motivated by new boundary term \eqref{eq:boundary} in the Chern-Simon formalism, a proper boundary term should be included in \eqref{eq:onshell} as well \footnote{The necessity of the new boundary terms explains the difference between the free energy of the thermal AdS with KdV boundary condition found in \cite{Dymarsky:2020tjh} and \cite{Erices:2019onl}.}.  The new boundary term can also be motived from the perspective of multi-trace deformation \cite{multitrace}. Only in the  Brown-Henneaux boundary condition which corresponds to
\bea 
\mathcal{H}_{BH}=\frac{1}{2\pi}\int d\varphi \, u,\quad f=1,
\eea  
these two masses are equal: $M_{\text{th}}=M_{\text{holo}}$. In general, $\mathcal{H}(u,u',\dots)$ can be viewed as a multi-trace deformation of $\mathcal{H}_{BH}$. For example, the simplest KdV boundary condition is described by
\be 
\mathcal{H}_{\text{KdV}}=\frac{1}{2\pi}\int d\varphi\, u+\frac{\lambda}{2\pi}\int d\varphi\, u^2\equiv Q_1+\lambda Q_3.
\ee 
The  KdV charge $Q_3$ of a two-dimensional conformal field theory is a double-trace operator
\be 
Q_3=\frac{1}{2\pi}\int d\varphi (TT),
\ee 
where $T$ is the holomorphic stress-energy tensor. Following the holographic dictionary of the double-trace deformation, the deformed action should be
\bea 
I_{Q_1+\lambda Q_3}=I_{Q_1}-\lambda \int dtd\varphi \frac{\delta Q_3}{\delta T}T.
\eea 
Next, we apply the general analysis to the three-dimensional C-metric.
\section{Application to C-metric}
To compare with the standard metric \eqref{eq:metric_static}, we first rewrite three-dimensional C-metric \eqref{eq:metric} in the FG gauge:
\be 
ds^2=\frac{1}{z^2}dz^2+\frac{1}{z^2}\(g^{(0)}+z^2 g^{(2)}+z^4 g^{(4)}\),
\ee
with
\be
g^{(0)}_{ij}dx^idx^j=-\omega(x)^2dt^2+\frac{\omega(x)^2dx^2}{(x^2-1)\Gamma^2},\quad \Gamma=1+a^2-a^2x^2,
\ee 
and
\bea 
&&g_{tt}=\frac{1+a^2}{2}-\frac{(x^2-1)\Gamma^2\omega'^2}{2\omega^2},\\
&&g_{x x}=\frac{1+a^2}{2(x^2-1)\Gamma^2}+\frac{x (1-3a^2(x^2-1))}{(x^2-1)\Gamma}\frac{\omega'}{\omega}+\frac{2\omega\omega''-3\omega'^2}{2\omega^2},\\
&&g^{(4)}=\frac{1}{4}g^{(2)}{g^{(0)}}^{-1}g^{(2)}.
\eea 
Therefore, one can make the following identifications
\be\label{eq:conformal}
f(\varphi)=\omega(x(\varphi)),\quad d\varphi=\frac{\omega dx}{\sqrt{x^2-1}\Gamma(x)},\quad u_0f_0^2=1+a^2.
\ee 
We start by showing that different choice conformal factors $\omega=f$ lead to the same orbit. Let $f_1({\varphi_1})$ and $f_2({\varphi_2})$ be two choices of conformal factors. The relation \eqref{eq:conformal} implies that $\varphi_1$ and $\varphi_2$ diffeomorphic to each other 
\be 
\frac{d\varphi_1}{f_1(\varphi_1)}=\frac{d\varphi_2}{f_2(\varphi_2)},
\ee 
so that $f_1,u_1$ and $f_2,u_2$ belong to the same co-adjoint orbit. 
Using the general results  immediately we obtain
\bea 
\frac{1}{f_0}&=&\frac{1}{2\pi}\int \frac{d\varphi}{f}=\frac{1}{2\pi}\int \frac{\sqrt{g^{(0)}_{xx}}dx}{\omega(x)}=\frac{L_{S^1}}{2\pi},\\
u_0&=&(1+a^2)\(\frac{L_{S^1}}{2\pi}\)^2,\quad T_{H}=\frac{\sqrt{1+a^2}}{2\pi},\\
\quad S_{BH}&=&\frac{c L_{S^1}}{6}\sqrt{1+a^2},\quad M_{\text{holo}}[\varphi]=\frac{c L_{S^1}}{24\pi}(1+a^2+\frac{1}{\pi}\oint d\varphi_0\{\varphi,\varphi_0\}), \label{eq:centropy}
\eea 
where $L_{S^1}$ is the proper length of the (asymptotic) $S^1$:
\be 
L_{S^1}=\frac{2 \coth ^{-1}\left(\frac{x_0}{\sqrt{\left(a^2+1\right) \left(x_0^2-1\right)}}\right)}{\sqrt{a^2+1}}.
\ee

\noindent\textbf{$M_{\text{th}}$ vs. $M_{\text{holo}}$.} To derive $M_{\text{th}}$, in practice, one should specify the boundary condition first and then solve $f$ and $u$ according to the equations of motion \eqref{eq:eom}. In the case of three-dimensional C-metric, when we choose a conformal factor $f$, the other parameter $u$ is then fixed by the identity \eqref{eq:cons} with $u_0f_0^2=1+a^2$. At once we know the functional relation between $f$ and $u$, we can use the definition of $f$ \eqref{eq:def_f} to solve a suitable $M_{\text{th}}$. For example, choosing  $f(\varphi_{\text{ADM}})$ to be the one appearing in the asymptotic boundary in the ADM formalism \cite{arnowitt1961coordinate,Arenas-Henriquez:2023hur} we find
\bea  
&&f(\varphi(\varphi_0))=\sqrt{\frac{2(1+a^2)}{2+a^2+a^2\cosh(2\sqrt{1+a^2}\varphi_0/f_0)}},\\
&&u=-2-a^2+3f^2, \quad \rightarrow \quad f=\sqrt{\frac{2+a^2+u}{3}},
\eea  
in our convention. Integrating \eqref{eq:def_f} gives
 the thermodynamics mass:
 \bea \label{eq:mth}
 M_{\text{th}}&=&\frac{c}{24\pi}\int \frac{2}{3\sqrt{3}}(2+a^2+u)^{3/2} d\varphi \nonumber\\
 &=&\frac{c}{24\pi}\int 2f^3d\varphi\neq M_{\text{holo}}[\varphi_{\text{ADM}}]. 
 \eea 

\begin{figure}
    \centering
     \includegraphics[scale=1]{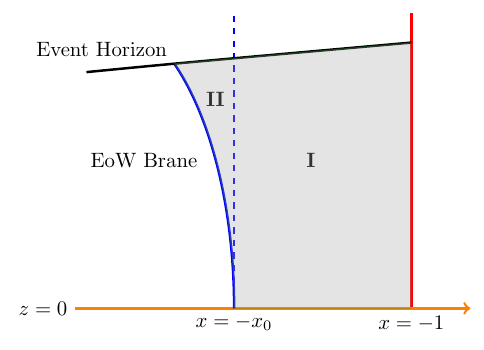}
    \caption{We have assumed that the metric \eqref{eq:metric_static} described two copies of the patch I. However, the EOW brane in the FG gauge is bended so the spacetime covered by the ADM gauge is I and II in the FG gauge.}
    \label{EOW brane in FG gauge}
\end{figure}

\noindent\textbf{ADM gauge vs. FG gauge.} 
It is demonstrated in \cite{Arenas-Henriquez:2023hur}, that the 3D C-metric within the ADM gauge exhibits certain anticipated holographic characteristics. For example, the Euclidean on-shell action equates to the black hole's free energy, and a conserved holographic stress-energy tensor can be explicitly specified. However, it is not clear how to formulate an appropriate first law of thermodynamics in ADM formalism even though that
the ADM mass is consistent with our result: $M_{\text{ADM}}=M_{\text{holo}}[\varphi_{\text{ADM}}]$.
Which gauge has the correct holographic description: the ADM gauge or the FG gauge? We expect that they are both correct. The difference lies in the fact that these two gauges cover different parts of the whole geometry as shown in Fig.\eqref{EOW brane in FG gauge}. In particular, the ADM gauge covers more horizon than the FG gauge does since in the ADM gauge the black hole entropy is larger than \eqref{eq:centropy} and the difference is :
\be 
S_{\text{ADM}}-S_{\text{FG}}=\frac{c}{3}\log\sqrt{\frac{1+a\sinh\psi_0}{1-a\sinh\psi_0}}=\frac{c}{3}\log\sqrt{\frac{1+\mu_0}{1-\mu_0}}=2g_{\mu_0},
\ee 
where $\mu_0$ is the tension of the EOW at $x=-x_0 (\text{or } \psi=-\psi_0)$. 
The extra piece $g_{\mu_0}$ exactly equals the boundary entropy of the EOW brane  implying that this extra entropy in the ADM gauge should be interpreted as the boundary entropy. The boundary entropy should not be included in the formulation of the first law of thermodynamics or we can add another term $-2T\delta  g_{\mu_0}$ in the first law. 
Below we  give a simpler example to confirm our interpretation.

\bigskip 

\noindent\textbf{A BTZ example.} The metric of the (non-rotating) BTZ black hole is 
\bea \label{BTZ}
ds^2=\frac{1}{z^2}\({f(z)} d \tau^2+\frac{\d z^2}{ f(z)}+ dx^2\),
\eea 
where  $f(z)=1-(z/z_H)^2,\tau\sim \tau+2\pi z_H$. In the BTZ geometry, the profile of the static EOW brane is \cite{Fujita:2011fp}
\bea 
Q_\pm(x_0,\lambda):\quad z= \pm \frac{z_H}{\lambda}\sinh \frac{x-x_0}{z_H},\quad \lambda=|\frac{\mu}{\sqrt{1-\mu^2}}|,\label{lambda}
\eea 
which ends on the AdS boundary at $x=x_0$. The tensionless brane is the constant $x$ surface. Let us consider a wedge $W_{0,\lambda}$ bounded by $x=0$ and a EOW brane $Q_+(x_0,\lambda)$ and a wedge $W_{0,0}$ bounded by $x=0$ and $x=x_0$. In the wedge $W_{0,0}$, the holographic mass, temperature and the entropy are
\bea 
T=\frac{1}{2\pi z_H},\quad S=\frac{c x_0}{6z_H},\quad M=\frac{c x_0}{24\pi z_H^2},
\eea 
which satisfies the first law
\be 
dM=-\frac{cx_0}{12\pi z_H^3}dz_H=TdS.
\ee 
However, for the wedge $W_{0,\lambda}$ we have
\be 
T=\frac{1}{2\pi z_H},\quad S_\lambda=\frac{cx_0}{6z_H}+\frac{c}{6}\sinh^{-1}\lambda,\quad M=\frac{c x_0}{24\pi z_H^2}.
\ee 
The difference of $S_\lambda$ and $S$ is exactly the boundary entropy corresponding to the horizon area difference in these two wedges. 
\section{Conclusions and Outlook}
In this work, we have described the thermodynamics and holographic properties of three-dimensional accelerating black holes in a more general and systematic way. In contrast to the previous discussion in the ADM formalism \cite{Arenas-Henriquez:2023hur}, our description is universal for all the choices of the conformal structure and with the help of the Chern-Simons formalism of the AdS gravity, the connection between the asymptotic boundary condition and the free energy can be built so that a proper first law of thermodynamics and the Smarr formula can be derived. The correct free energy is the starting point to study its phase transition, for example, the Hawking-Page phase transition \cite{HP}. The KdV black holes have a very fruitful phase structure \cite{Dymarsky:2020tjh, Wang}, and we expect a similar phenomenon will happen in three-dimensional accelerating black holes. 

The static solution \eqref{eq:metric_static} can be generalized to include angular momentum very easily by exciting non-equal left $u,f$ and right $\bar{u},\bar{f}$ movers. Then it is very promising to construct an accelerating black hole with rotation in the FG gauge. The free energy will be simply given by $\mathcal{F}+\bar{\mathcal{F}}$. The challenge is to rewrite the solution in the ``intrinsic" gauge like \eqref{eq:metric} because the coordinate transformation between \eqref{eq:metric} and the FG gauge is only known as an infinite series expansion of $z$.  It is also interesting to generalize our approach to describe the charged accelerating BTZ black holes \cite{EslamPanah:2022ihg}.

On the holographic property, it is important to study the entanglement entropy. As we have shown the bulk theory is not just the AdS Einstein gravity in general. It is possible that RT formula $S_A=\frac{\text{Area}_{A}}{4G_N}$ is modified. Moreover, the EOW branes also serve as  RT surface anchors which may lead to a phase transition of entanglement entropy. We leave these interesting questions in the future work.

Finally, it is worth revisiting Class I and Class III type solutions in our approach. In particular, the Class I$_C$ solution describes an accelerating black hole solution parametrically disconnected from the standard BTZ geometry. The difference between the Class I$_C$ solution and the Class II solution may be more transparent in our approach.
\section*{Acknowledgments}
We thank Huajia Wang, Cheng Peng, and Nozaki Masahiro for the valuable discussion and collaboration on related topics. JT is supported by  the National Youth Fund No.12105289.
\printbibliography
\end{document}